# Absorbed dose evaluation of Auger electron-emitting radionuclides: impact of input decay spectra on dose point kernels and *S*-values

[a]Nadia Falzone[1,2*], [a]Boon Q. Lee[3], José M. Fernández-Varea[4], Christiana Kartsonaki[5], Andrew E. Stuchbery[3], [b]Tibor Kibédi[3], [b]Katherine A. Vallis[1]

[1]CR-UK/MRC Oxford Institute for Radiation Oncology, Department of Oncology, University of Oxford, Oxford, United Kingdom

[2]Department of Biomedical Science, Tshwane University of Technology, Pretoria, South Africa

[3]Department of Nuclear Physics, Research School of Physics and Engineering, Australian National University, Canberra, ACT 2601, Australia

[4]Facultat de Física (FQA and ICC), Universitat de Barcelona, Diagonal 645, E-08028 Barcelona, Spain

[5]Nuffield Department of Population Health, University of Oxford, Oxford, United Kingdom

[a]Joint first authors, [b]Joint last authors

**Running title:** Dose point kernels and *S*-values of Auger electron emitters

[*]Corresponding author:

Dr. Nadia Falzone

Department of Oncology | University of Oxford

CRUK/MRC Oxford Institute for Radiation Oncology,

Old Road Campus Research Building, Off Roosevelt Drive. Oxford. OX3 7DQ

T: +44 (0)1865 225841  F: +44 (0) 1865 857127

E-mail: nadia.falzone@oncology.ox.ac.uk

## ABSTRACT

**Aim:** The aim of this study was to investigate the impact of decay data provided by the newly developed stochastic atomic relaxation model *BrIccEmis* on dose point kernels (DPKs- radial dose distribution around a unit point source) and *S*-values (absorbed dose per unit cumulated activity) of 14 Auger electron (AE) emitting radionuclides, namely $^{67}$Ga, $^{80m}$Br, $^{89}$Zr, $^{90}$Nb, $^{99m}$Tc, $^{111}$In, $^{117m}$Sn, $^{119}$Sb, $^{123}$I, $^{124}$I, $^{125}$I, $^{135}$La, $^{195m}$Pt and $^{201}$Tl.

**Methods:** Radiation spectra were based on the nuclear decay data from the Medical Internal Radiation Dose (MIRD) RADTABS program and the *BrIccEmis* code, assuming both an isolated-atom and condensed-phase approach. DPKs were simulated with the PENELOPE Monte Carlo (MC) code using event-by-event electron and photon transport. *S*-values for concentric spherical cells of various sizes were derived from these DPKs using appropriate geometric reduction factors.

**Results:** The number of Auger and Coster Kronig (CK) electrons and X-ray photons released per nuclear decay (yield) from MIRD-RADTABS were consistently higher than those calculated using *BrIccEmis*. DPKs for the electron spectra from *BrIccEmis* were considerably different from MIRD-RADTABS in the first few hundred nanometres from a point source where most of the Auger electrons are stopped. *S*-values were, however, not significantly impacted as the differences in DPKs in the sub-micrometre dimension were quickly diminished in larger dimensions.

**Conclusion:** Overestimation in the total AE energy output by MIRD-RADTABS leads to higher predicted energy deposition by AE emitting radionuclides, especially in the immediate vicinity of the decaying radionuclides. This should be taken into account when MIRD-RADTABS data are used to simulate biological damage at nanoscale dimensions.

1. **Introduction**

When evaluating the efficacy of novel radionuclide constructs used in diagnostic and therapeutic nuclear medicine, it is often necessary to estimate the absorbed dose from intracellularly localised radionuclides at a single-cell level. To enable these calculations, the Medical Internal Radiation Dose (MIRD) Committee (Goddu *et al.*, 1997) has compiled comprehensive tables of cellular *S*-values (absorbed dose per unit cumulated activity). An important aspect of cellular *S*-value calculations is the adopted decay radiation spectrum. The radiation spectra used in the MIRD compilation of *S*-values (Goddu *et al.*, 1997) were taken from the work of Eckerman et al. (Eckerman *et al.*, 1993), which is based on the nuclear structure data from the Evaluated Nuclear Structure Data Files (ENSDF). More recently, MIRD (Eckerman and Endo, 2008a) published up-to-date tabulations of radiation spectra for 333 radionuclides. In the calculation of atomic relaxation spectra emitted after electron capture (EC) or internal conversion (IC), several assumptions were made in the MIRD tabulations partly due to computational limits at the time and the paucity of experimental data. These include the use of binding energies of neutral atoms and neglect of spectator vacancies that could potentially lead to inaccurate estimates of the emissions of low-energy X-rays and AE. An improved computational route is therefore required to address the limitations of the MIRD committee data as suggested by the International Nuclear Data Committee of the International Atomic Energy Agency (Nichols *et al.*, 2011).

The nuclear structure data from ENSDF is used to evaluate the distribution of initial vacancies, which determines the initial atomic state before the cascade simulation commences. Furthermore, two approaches toward the fate of valence vacancies produced during the cascade have been considered in the literature, namely the isolated-atom and condensed-phase conditions (Howell, 1992; Pomplun, 2000; Stepanek, 2000). The distinction between these two approaches is the assumption regarding the neutralisation of any valence vacancy during a cascade. In the condensed-phase approach, charge transfer between the environment and the valence shell is assumed to take place during the cascade, leaving the atom completely neutralised at the end of the cascade process, while with the isolated-atom approach, the atom becomes ionised. Detailed comparison of isolated-atom and condensed-phase approximations and the impact of partial neutralisation have been discussed by Charlton et al (Charlton *et al.*, 1983). There is no consensus in the literature as to which approach should be adopted, due to the lack of experimental evidence for the time-scale of electron transfer from DNA, proteins or a liquid water environment to an Auger-emitter. For examples, radiation spectra in the MIRD tabulations (Eckerman and Endo, 2008a) and Howell (Howell, 1992) were calculated using the condensed-phase approach, while Pomplun (Pomplun, 2000) adopted the isolated-atom approximation. This is one of the main factors contributing to the variability in published emission spectra of selected Auger-emitters (Howell, 1992; Pomplun, 2000; Stepanek, 2000; Eckerman and Endo, 2008a). Recently Lee and co-workers (Lee *et al.*, 2012; Lee *et al.*, 2016) published a stochastic atomic relaxation cascade (referred to as vacancy cascade) model called *BrIccEmis* to simulate the AE and X-ray spectra of elements up to Fermium, $Z=100$. In the present work, emission spectra based on both approaches were calculated using *BrIccEmis* for detailed comparison.

The main aim of this study is therefore to investigate the impact of new decay data provided by *BrIccEmis* (Lee *et al.*, 2012; Lee *et al.*, 2016) on dose point kernels (DPKs) and *S*-values of 14 Auger electron-emitting radionuclides[1], namely $^{67}$Ga (EC), $^{80m}$Br (IT), $^{89}$Zr (EC $\beta^+$), $^{90}$Nb (EC $\beta^+$), $^{99m}$Tc (IT $\beta^-$), $^{111}$In (EC), $^{117m}$Sn (IT), $^{119}$Sb (EC), $^{123}$I (EC), $^{124}$I (EC $\beta^+$), $^{125}$I (EC), $^{135}$La (EC), $^{195m}$Pt (IT) and $^{201}$Tl (EC) (Eckerman and Endo, 2008a).

## 2. Materials and methods

---

[1] Decay mode: $\beta^-$ beta minus, $\beta^+$ positron emission/beta plus, EC electron capture, and IT internal transition.

*2.1. Stochastic cascade model*

A detailed description of *BrIccEmis* has been published elsewhere (Lee *et al.*, 2012; Lee *et al.*, 2016). The initial vacancy distributions due to EC and internal conversion (IC) of γ-emission of 14 radionuclides were evaluated using the nuclear structure data in the ENSDF format, available from www.nndc.bnl.gov/ensdf/. The energies of all possible radiative and non-radiative atomic transitions to fill a vacancy were checked during the cascade and any energetically forbidden transitions were discarded before a propagation step. A transition energy was set equal to the difference between the total energies of the atom before and after a transition, calculated using the RAINE code (Band *et al.*, 2002), which is based on the relativistic Dirac-Fock self-consistent method. The atomic transition probabilities were taken from the Evaluated Atomic Data Library (EADL) (Perkins *et al.*, 1991). Since all probabilities from EADL were calculated for singly-ionised systems, the empirical correction of Krause and Carlson (Krause and Carlson, 1967) was applied to modify probabilities according to the actual number of electrons available in the participating atomic subshells. Radiation spectra are provided as supplementary data (Supplementary Figures 1-7).

*2.2. Dose point kernels*

Electron and X-ray transport was performed by event-by-event Monte Carlo (MC) simulations with the general-purpose MC code PENELOPE (Salvat *et al.*, 2011). MC transport of the complete radiation spectra based on the unabridged decay data from MIRD tabulation (MIRD-RADTABS software Ver. 2.2) (Eckerman and Endo, 2008a), as well as the unabridged spectra generated by *BrIccEmis* based on the isolated-atom and condensed-phase approximations were used for all radionuclides considered. In the case of $^{90}$Nb, $^{117m}$Sn, and $^{124}$I, the abridged AE+CK+IC spectra were utilised as their unabridged spectra were not provided in MIRD-RADTABS. Since $^{135}$La was not included in MIRD-RADTABS, the abridged AE+CK+IC spectrum of $^{135}$La was adopted from the International Commission on Radiological Protection (ICRP) Publication 107 (Eckerman and Endo, 2008b).

An isotropic point source was placed in an infinite liquid water medium (mass density $\rho$ = 1 g/cm$^3$) and the mean absorbed dose from the emitted ionising radiation (AE, CK, IC, β$^+$particles and X-rays) was recorded in 1-nm-thick spherical shells around the decay site. Mean absorbed doses were tallied up to a radial distance of 50 µm from the point source. Electron and photon cut-off energies were set at 50 eV. Typically 2x10$^9$ primary particles were simulated in each run.

*2.3. S-values from simulated DPKs*

Cell ($R_C$) and nucleus ($R_N$) radii combinations for concentric spherical cells as tabulated by MIRD (Goddu *et al.*, 1997) were expanded to include larger cell sizes (up to $R_C$ = 12 µm and $R_N$ = 11 µm). *S*-values (Gy/Bq·s) were calculated as previously described (Falzone *et al.*, 2015). Briefly, taking the cell nucleus as the target, *S*-values were calculated from the simulated DPKs to determine contributions from the nucleus (N←N), the cytoplasm (N←Cy) or the cell surface (N←CS) as;

$$S(\text{T} \leftarrow \text{S}) = \frac{1}{m_\text{T}} \int_0^\infty \mathrm{d}r\, 4\pi r^2 \rho\, \psi_{\text{T}\leftarrow\text{S}}(r) D(r),$$

(1)

where the DPK $D(r)$ is the mean absorbed dose scored in a 1-nm-thick spherical shell at a distance *r* from the point source, $\rho$ is the mass density, $m_\text{T}$ is the mass of the target volume and $\psi_{\text{T}\leftarrow\text{S}}(r)$ are the geometric reduction factors. $\psi_{\text{T}\leftarrow\text{S}}(r)$ denotes the probability of a randomly directed vector of length *r* that starts from a random point within the source region (S) and ends within the target region (T) (Goddu *et al.*, 1997); these factors become analytical functions in the case of concentric nucleus and cell configurations.

Tabulated MIRD *S*-values (Goddu *et al.*, 1997) were calculated with DPKs assuming straight line trajectories and the continuous slowing down approximation. To enable a direct comparison of *S*-values derived from MC DPKs, thus just taking the differences in spectra (Eckerman and Endo 2008a,b) vs. *BrIccEmis* and not calculation method into account, we compared *S*-values derived from event-by-event MC simulations as previously published (Falzone et al 2015) with the *BrIccEmis* derived *S*-values.

## 3. Results

*3.1. AE and X-ray yields*

A summary of AE, CK, and X-ray yields, where yield is defined as the number of a particular radiation type released per nuclear decay, comparing MIRD-RADTABS (Eckerman and Endo, 2008a) and *BrIccEmis* data is provided in Table 1. AE and CK yields calculated using *BrIccEmis* are generally lower than the values provided by MIRD-RADTABS even though the condensed-phase approximation is in better agreement with MIRD-RADTABS for several radionuclides. Most of the discrepancies come from the differences in yields of outer-shell non-radiative transitions (M-shell and above). There are significant differences between the condensed-phase approximation and MIRD-RADTABS data for $^{117m}$Sn and $^{119}$Sb. In the case of $^{117m}$Sn, the main source of the discrepancy is an unassigned non-radiative transition in Sn at 11.33 eV with a yield of 7.6. For $^{119}$Sb, in which the vacancy cascade occurs in the atomic system of Sn following EC decay, MIRD-RADTABS data included 9.3 non-radiative transitions that involve the O3-subshell which is only occupied in the excited atomic states of Sn. *BrIccEmis* ignored these transitions because the probability of finding a Sn atom in its excited state at human body temperature is extremely low (< 0.001) according to the Boltzmann distribution (Kramida *et al.*, 2015).

Both condensed-phase and isolated-atom approximations result in similar X-ray yields. However, both approximations lead to significantly lower X-ray yields than MIRD-RADTABS data because, in the latter, energy released during the neutralisation of a valence vacancy is assigned to radiative recombination, thus producing a very high number of the so-called "free-bound low-energy X-rays". This approach could be misleading as femtosecond spectroscopy has revealed that vibrational relaxation is competing with electronic relaxation in electron transfer at a molecular level within the timescale of the vacancy cascade (Anderson and Lian, 2004; Frischkorn and Wolf, 2006). The correct treatment of the neutralisation of valence vacancies, which is dependent on the environment, is currently beyond the scope of *BrIccEmis*. Hence, the code only reports bound-bound radiative transitions.

*3.2. Dose point kernels*

**Electrons** – DPKs of $^{111}$In, $^{124}$I, $^{125}$I and $^{201}$Tl derived from MIRD-RADTABS data (Eckerman and Endo, 2008a) are shown in the upper panel of Figure 1 in which the contributions from AE, CK, IC, and $\beta^+$ are included. The shoulders in the DPKs are a result of the energy deposition contributions of the aforementioned particulate radiations. The DPKs derived from *BrIccEmis* data, which are given as ratios to the DPKs calculated using MIRD-RADTABS data, are shown in the lower panels of Figure 1. DPKs of the remaining radionuclides are supplied as supplementary Figures (Supplementary Figures 8-10).

In general, little discrepancy exists between the condensed-phase and isolated-atom approximations at large radii. The two deviate at small radii as this is the domain where low-energy AE and CK electrons, which are emitted from M and higher shells towards the final stage of the vacancy cascade, deposit their energies. The radial interval where the two approaches deviate from each other is radionuclide dependent; the deviation begins at about 20 nm and 300 nm in $^{111}$In and $^{201}$Tl, respectively (Figure 1 a and d). DPKs derived from the *BrIccEmis* data differ from DPKs calculated using MIRD-RADTABS data up to the maximum range of L-Auger electrons. At large radii, DPKs derived from all three approaches are alike for $^{111}$In and $^{124}$I. For $^{125}$I and $^{201}$Tl (Figure 1 c and d), *BrIccEmis*-based DPKs are dissimilar from MIRD-based DPKs due to the differences in IC yields. Besides, many low-energy AE and CK electrons are stopped at the decay site due to the cut-off energy of 50 eV in electron transport, resulting in the discontinuities in DPKs.

There are noticeable oscillations in the DPK plots (lower panels). These are due to the differences between MIRD-RADTABS and *BrIccEmis* AE and CK electron energies emitted from the L-shell and above. Figure 1(a) shows that *BrIccEmis*-based DPKs are slightly higher than MIRD-based DPKs near the maximum range of the L-Auger electrons. The latter becomes dominant at smaller radii around the median range of the L-Auger group before the former overtakes it again near the maximum range of the M-Auger group. This is a result of an energy shift in the L-Auger spectrum from MIRD-RADTABS. Figure 2 compares spectra from *BrIccEmis* (condensed-phase and isolated-atom approximations are the same for L-Auger) and MIRD-RADTABS with experimental data (Yakushev *et al.*, 2005). The comparison shows an energy shift towards the higher energies in MIRD-RADTABS data while *BrIccEmis* data are in good agreement with the experiment within the given energy range. Similar behaviour is found at smaller radii where AE emitted from higher atomic shells are dominant. This is prevalent in all other radionuclides except $^{90}$Nb, $^{117m}$Sn, $^{124}$I, and $^{135}$La in which the abridged spectra from MIRD-RADTABS were used (Figure 1(b) and Supplementary Figures 8(b); 9(b) and 10(a)).

In cases where MIRD-RADTABS's abridged spectra ($^{90}$Nb, $^{117m}$Sn and $^{124}$I), which are sometimes called "average spectra" were used, *BrIccEmis*-based DPKs are significantly higher than MIRD-based DPKs near the maximum range of an Auger group. Around the median range of a group, the latter surpasses the former. Since transition intensity is concentrated at the mean energy of a group in the abridged spectra, *BrIccEmis*-based DPKs eventually overtake MIRD-based DPKs at smaller radii.

**X-rays** – DPKs of $^{125}$I and $^{201}$Tl for all three X-ray spectra (*BrIccEmis*' condensed-phase and isolated-atom assumptions and MIRD-RADTABS) are depicted in Figure 3. The DPKs derived from the isolated-atom approximation are significantly higher than the other two DPKs over the range 2 nm – 1 µm for both radionuclides. This is a consequence of the higher yields of low-energy X-rays (25 - 500 eV) from the isolated-atom approximation (Supplementary Figures 1-7). The contributions from the MIRD-RADTABS X-ray spectra, for distances smaller than 2 nm from the point source, exceed those of the X-ray spectra from the *BrIccEmis* code. The so-called free-bound X-rays reported in MIRD-RADTABS data are very low in energy (< 50 eV). Thus, they were absorbed at the site of decay in the PENELOPE simulations resulting in the spikes in X-ray DPKs. This is the same in the X-ray DPKs derived for all other radionuclides (Supplementary Figures 11-13).

*3.3. Cellular S-values*

Comparisons of *S*-values derived from MIRD-RADTABS data (Eckerman and Endo, 2008a) as previously published (Falzone *et al.*, 2015) with those presently calculated from the *BrIccEmis* spectra are shown in supplementary Tables 1-6 (except $^{124}$I and $^{135}$La, which were not reported in the previous publication). MC calculated *S*-values from *BrIccEmis* data are generally in good agreement with the values derived from MIRD-RADTABS data, even though the difference between the data is significant in the immediate vicinity of the decay site (see Figure 1), the calculations of *S*-values at a cellular scale diminish the contributions of nanoscale dose where Auger electrons dominate. For low atomic number radionuclides ($^{67}$Ga and $^{80m}$Br), differences are less than 1% for *S*(N←CS) and *S*(N←Cy) values and less than 3% for *S*(N←N) values (Supplementary Table 1). However, with increasing atomic number, the variation in *S*(N←N) contributions from the various spectra data increase markedly, with the isolated-atom assumption contributing to differences up to 18% (for $^{201}$Tl) for small $R_C$ and $R_N$

(Supplementary Table 6) while the condensed-phase assumption differs by less than 10% from MIRD-RADTABS data for all radionuclides (Supplementary Tables 1-6). In summary, a decrease in *S*-values is more prominent for small cell sizes, as the differences between MIRD-RADTABS and *BrIccEmis* data are most noticeable for electrons emitted from the L-shell and above which usually have a maximum range of less than 1 µm. This point is further illustrated in Figure 4, dose point kernels of MIRD-RADTABS and *BrIccEmis* data can be significantly different at close range from the source whereas *S*-values remain similar at the cell level. Over shorter dimensions at a DNA scale, MIRD-RADTABS data would overestimate the calculated dose as compared to *BrIccEmis* data.

As *S*-values have not been supplied for $^{135}$La in the MIRD monograph of cellular *S*-values (Goddu *et al.*, 1997), we have included these calculations for both the isolated-atom and condensed-phase assumptions (Table 2). IC electrons contributed at most 0.02% when $R_N$ was large relative to $R_C$ in $^{135}$La. Similarly, *S*-values of $^{124}$I not included in our previous work (Falzone *et al.*, 2015), are now derived from *BrIccEmis* data and are shown in supplementary Table 7. β$^+$ particles contribute up to 12% of the *S*-values while IC, similar to the case of $^{135}$La, contribute at most 0.18% when $R_N$ is large relative to $R_C$.

Where available, a comparison of *S*-values ($R_C$=5 µm, $R_N$=4 µm) with those derived from other computational codes in the literature is also presented (Table 3). For overlapping source and target configurations, *BrIccEmis* data compare well with MC4 (Bousis *et al.*, 2010), ETRACK (Ftániková and Böhm, 2000) and GEANT4 (Šefl *et al.*, 2015) data, with differences < 10%, apart from $^{201}$Tl. However, differences greater than 40% between *BrIccEmis* and ETRACK data are seen for *S*(N←CS) configurations. All aforementioned publications adopted the average spectra from Howell (Howell, 1992) which are similar to the abridged spectra of MIRD-RADTABS. From the comparisons of *S*-values derived from *BrIccEmis* and MIRD-RADTABS for $^{90}$Nb and $^{117m}$Sn, differences due to input decay spectra are expected not to be more than 10%. Thus the large discrepancies between *BrIccEmis* and ETRACK are mainly due to the differences in the MC transport codes rather than radiation spectra.

The contributions of X- and γ-rays to the *S*-values were ignored in the MIRD Cellular *S*-values monograph (Goddu *et al.*, 1997) as they reportedly contributed less than 2% to the total *S*-value $S_{TOT} = S(N \leftarrow N) + S(N \leftarrow Cy) + S(N \leftarrow CS)$. However, for very small $R_C$ and $R_N$ and for increasing atomic numbers, the X-ray spectra taken from MIRD-RADTABS data (Eckerman and Endo, 2008a) may contribute as much as 5% to the total *S*-values calculated from equation (1) (Supplementary Table 8). In contrast, the total contribution of X-rays to *S*-value calculations from *BrIccEmis* is less than 1% for all radionuclides considered (Supplementary Table 9 – data presented for isolated-atom assumption only).

4. **Discussion**

Energies of X-rays, AE and CK electrons in the spectra from MIRD-RADTABS (Eckerman and Endo, 2008a) were extracted directly from EADL (Perkins *et al.*, 1991) which is based on singly-ionised atomic systems. This simplification, which was implemented partly due to computational limitations at the time, neglects the variations in transition energies caused by the presence of spectator vacancies during the vacancy cascade. Previous findings of cascade modelling agreed that each atomic transition energy should be calculated from the electronic configurations involved in that particular event (Howell, 1992; Pomplun, 2000). Even in the case of the L-Auger group, where the effect of spectator vacancies is expected to be minimal, the spectrum derived from this method exhibits a shift towards higher energies (see Figure 2).

Most non-radiative transitions, especially those involving higher atomic subshells, have lower energies than the values reported in EADL and some transitions simply become energetically forbidden after the effects of spectator vacancies are properly accounted for. Consequently, *BrIccEmis* spectra yielded lower energy deposition than MIRD-RADTABS spectra within the first 100 nm, where most of the AE and CK electrons are stopped (Figure 1 and Supplementary Figures 8-10). Experimental spectra of low-energy AE and CK electrons (< 1200 eV) from radioactive decays are extremely scarce in the literature. One such energy spectrum (Kovalík *et al.*, 2004) is highly distorted thus useful comparisons could not be made. Future measurements of low-energy electrons with cleaner background are desired for more detailed comparison between theory and experiment.

The isolated-atom approximation produced more low-energy X-rays (bound-bound transitions) than the condensed-phase approximation (Supplementary Figures 1-7) owing to the inhibition of low-energy AE and CK electrons near the end of the vacancy cascade. The so-called free-bound X-rays from MIRD-RADTABS could increase the total contribution of X-rays to the *S*-values up to 5% at small cell sizes (see subsection Cellular *S*-values). Nevertheless, simulations of photon transport below the current cut-off energy (50 eV) would resolve the sharp fall seen in X-ray DPKs (Figure 3 and Supplementary Figures 11-13) derived from MIRD-RADTABS and reduce the total contribution of X-rays to the *S*-values.

Notwithstanding the fact that vibrational relaxation is a competing process with electronic relaxation, the mechanism of valence-vacancy neutralisation in a molecular environment is further complicated by other processes that could emit ultra-low energy electrons (ULEE) rather than photons, such as interatomic Coulombic decay (Cederbaum *et al.*, 1997; Stumpf *et al.*, 2016). These processes, which occur as a terminal decay step of vacancy cascades, could emit extra ULEE which would otherwise be energetically forbidden in an atomic environment. Given that the cut-off energy in simulations was set at 50 eV, the emission of these electrons could affect the presently calculated energy deposition only in the first few nanometres from the source. These processes are environment-dependent processes and they are beyond the scope of the current version of *BrIccEmis*.

One of the main factors contributing to the large variation of calculated AE and CK yields reported in literature, is the assumption on whether valence-vacancy neutralisation happens during the vacancy cascade which has a very short lifetime (~ $10^{-16} - 10^{-13}$ s). The condensed-phase approximation differs from the isolated-atom approach at radii below the maximum range of M-shell Auger electrons (Figure 1 and Supplementary Figure 8-10). These differences are mainly found within the first 100 nm except in heavy atomic systems such as those involved in the decay of $^{195m}$Pt and $^{201}$Tl. Recent ultrafast spectroscopic studies have shown that the electron transfer at certain molecule-nanoparticle interfaces is within a few hundreds of femtoseconds (Anderson and Lian, 2004; Furube *et al.*, 2014). These experiments indicate that the electron transfer rate could be as fast as the vacancy cascade in some circumstances. The reported electron transfer rate is, however, dependent on the properties of the compounds as well as the environment. Better designed experiments using radiopharmaceutical drugs could be crucial in resolving the aforementioned ambiguity. Detection of very low-energy X-rays could be used as an alternative method to distinguish the two approximations since the isolated-atom approximation gives rise to measureable amount of X-rays with energies below 500 eV.

The use of abridged decay data is common in the calculation of *S*-values in the literature (Bousis *et al.*, 2010; Ftániková and Böhm, 2000; Šefl *et al.*, 2015). Our work shows that the use of averaged data is not critical for the calculation of cellular *S*-values. Nevertheless, unabridged decay data is better as input for the study of sub-micrometre energy deposition since the abridged version could artificially change the dose distribution at this scale.

Although DPKs are shown to be highly sensitive to the differences in input decay spectra, cellular *S*-values are, however, relatively unaffected except for heavy atomic systems within the isolated-atom approach. Differences in the first 100 nm between the DPKs derived from MIRD-RADTABS and *BrIccEmis* spectra are rapidly diminished for larger dimensions. This is evident from a comparison of the energy deposition as a function of DNA dimensions. Figure 4 compares the energy deposition based on spectra from MIRD-RADTABS with *BrIccEmis* in spheres of diameters representing different DNA condensation states. The absorbed doses derived from MIRD-RADTABS spectra are significantly higher in the first 30-nm diameters. However, these differences are considerably reduced as diameters representing DNA condensation states increase.

Taking into consideration that the potential radiation risk of AE emitters used in diagnostic nuclear medicine has been highlighted, especially in the case of intra-nuclear internalisation of the radiopharmaceutical (Howell 2011), dose calculations using MIRD-RADTABS data (Eckerman and Endo 2008) could potentially lead to an overestimation of radiation risk at sub-cellular dimensions. In contrast, when nuclear targeting with an AE emitter is desired for therapy it is essential to maximise the absorbed dose, thus necessitating the use of accurate decay spectra for dose calculation. Geant4, which has been used to assess the dose enhancement of high-*Z* nanoparticles in radiotherapy and proton therapy (Butterworth *et al.*, 2012; Yuting *et al.*, 2014), has recently incorporated a MIRD-RADTABS-esque atomic vacancy cascade approach to generate AE and CK energy spectra for investigations into applications of these nanoparticles in medicine (Incerti *et al.*, 2016). The dose enhancement in the first few hundred nanometres could be affected by the accuracy of AE and CK energy spectra. In the future, the effect of different AE and CK energy spectra on the dose enhancement of high-Z nanoparticles should be carefully investigated.

## 5. Conclusion

We conclude that the choice of atomic relaxation spectrum after nuclear decay (*BrIccEmis* vs. MIRD-RADTABS (Eckerman and Endo, 2008a) and *BrIccEmis*-isolated vs *BrIccEmis*-condensed) considerably affects the DPKs of AE emitting radionuclides up to several hundred nanometres. This implies that caution should be employed when evaluating the energy deposition of an AE emitting radionuclide over DNA dimensions. Nevertheless, these differences did not significantly affect *S*-values which are calculated for volumes of larger dimension as most AE and CK electrons deposit their energies within the first micron from the decay site. Although the two different approaches concerning the fate of valence vacancies produced during the vacancy cascade (isolated vs condensed) yield significantly different numbers of AE and CK electrons, the discrepancy in DPKs due to different electron yields is mainly restricted to the first 100 nm from a point source.

**Acknowledgments**

The authors gratefully acknowledge funding support from the Cancer Research-UK (C5255/A15935), the Medical Research Council (MC_PC_12004), the Australian Research Council Discovery Grant (no. DP140103317) and the Generalitat de Catalunya (project no. 2014 SGR 846).

**Table 1.** Comparison of AE+CK and X-ray yields from *BrIccEmis* and MIRD-RADTABS (Eckerman and Endo, 2008a).

| Nuclide | Yield of AE+CK | | | Yield of X-rays | | |
|---|---|---|---|---|---|---|
| | *BrIccEmis* Isolated | *BrIccEmis* Condensed | MIRD | *BrIccEmis* Isolated | *BrIccEmis* Condensed | MIRD |
| $^{67}$Ga | 4.56 | 4.85 | 4.96 | 0.57 | 0.57 | 6.87 |
| $^{80m}$Br | 7.33 | 8.96 | 9.60 | 0.82 | 0.81 | 12.0 |
| $^{89}$Zr | 4.58 | 6.88 | 9.45 | 0.51 | 0.51 | 10.7 |
| $^{90}$Nb | 4.33 | 6.88 | 8.77 | 0.52 | 0.52 | 10.1 |
| $^{99m}$Tc | 3.52 | 4.39 | 4.41 | 0.08 | 0.08 | 5.58 |
| $^{111}$In | 5.84 | 7.17 | 7.43 | 0.91 | 0.91 | 9.50 |
| $^{117m}$Sn | 5.38 | 5.72 | 14.2 | 0.80 | 0.76 | 16.1 |
| $^{119}$Sb | 10.0 | 14.4 | 23.7 | 0.97 | 0.88 | 26.4 |
| $^{123}$I | 7.38 | 12.3 | 13.7 | 1.09 | 0.97 | 15.8 |
| $^{124}$I | 5.04 | 8.38 | 9.17 | 0.72 | 0.65 | 10.6 |
| $^{125}$I | 11.9 | 20.0 | 23.0 | 1.76 | 1.57 | 26.5 |
| $^{135}$La | 7.44 | 10.9 | 12.3† | 0.93 | 0.81 | 14.3† |
| $^{195m}$Pt | 20.4 | 30.6 | 36.6 | 1.21 | 1.20 | 41.1 |
| $^{201}$Tl | 10.9 | 18.8 | 20.9 | 1.53 | 1.45 | 24.3 |

†ICRP Publication 107 (Eckerman and Endo, 2008b).

**Table 2**. Cellular $S$-values (Gy/Bq·s) for $^{135}$La, calculated both with the condensed-phase and isolated-atom assumptions of the *BrIccEmis* code.

| $R_N$ (μm) | $R_C$ (μm) | Condensed-phase spectra | | | Isolated atom spectra | | |
|---|---|---|---|---|---|---|---|
| | | $S(N\leftarrow N)$ | $S(N\leftarrow Cy)$ | $S(N\leftarrow CS)$ | $S(N\leftarrow N)$ | $S(N\leftarrow Cy)$ | $S(N\leftarrow CS)$ |
| 1 | 3 | 1.40E-01 | 1.11E-03 | 1.65E-04 | 1.33E-01 | 1.10E-03 | 1.62E-04 |
| 2 | 3 | 1.91E-02 | 7.97E-04 | 1.76E-04 | 1.83E-02 | 7.88E-04 | 1.72E-04 |
| 2 | 4 | 1.91E-02 | 3.59E-04 | 1.09E-04 | 1.83E-02 | 3.54E-04 | 1.07E-04 |
| 3 | 4 | 5.92E-03 | 3.36E-04 | 1.16E-04 | 5.66E-03 | 3.32E-04 | 1.13E-04 |
| 2 | 5 | 1.91E-02 | 2.20E-04 | 7.99E-05 | 1.83E-02 | 2.16E-04 | 7.82E-05 |
| 3 | 5 | 5.92E-03 | 1.86E-04 | 8.17E-05 | 5.66E-03 | 1.83E-04 | 8.00E-05 |
| 4 | 5 | 2.58E-03 | 1.89E-04 | 8.53E-05 | 2.47E-03 | 1.86E-04 | 8.35E-05 |
| 3 | 6 | 5.92E-03 | 1.31E-04 | 6.36E-05 | 5.66E-03 | 1.29E-04 | 6.22E-05 |
| 4 | 6 | 2.58E-03 | 1.20E-04 | 6.42E-05 | 2.47E-03 | 1.17E-04 | 6.28E-05 |
| 5 | 6 | 1.37E-03 | 1.23E-04 | 6.58E-05 | 1.31E-03 | 1.21E-04 | 6.44E-05 |
| 3 | 7 | 5.92E-03 | 1.01E-04 | 5.14E-05 | 5.66E-03 | 9.94E-05 | 5.03E-05 |
| 4 | 7 | 2.58E-03 | 9.10E-05 | 5.10E-05 | 2.47E-03 | 8.93E-05 | 4.99E-05 |
| 5 | 7 | 1.37E-03 | 8.49E-05 | 5.08E-05 | 1.31E-03 | 8.33E-05 | 4.97E-05 |
| 6 | 7 | 8.16E-04 | 8.69E-05 | 5.13E-05 | 7.82E-04 | 8.54E-05 | 5.02E-05 |
| 4 | 8 | 2.58E-03 | 7.39E-05 | 4.10E-05 | 2.47E-03 | 7.25E-05 | 4.01E-05 |
| 5 | 8 | 1.37E-03 | 6.75E-05 | 4.02E-05 | 1.31E-03 | 6.62E-05 | 3.94E-05 |
| 6 | 8 | 8.16E-04 | 6.29E-05 | 3.97E-05 | 7.82E-04 | 6.17E-05 | 3.89E-05 |
| 7 | 8 | 5.30E-04 | 6.37E-05 | 3.99E-05 | 5.08E-04 | 6.26E-05 | 3.90E-05 |
| 5 | 9 | 1.37E-03 | 5.61E-05 | 3.18E-05 | 1.31E-03 | 5.50E-05 | 3.12E-05 |
| 6 | 9 | 8.16E-04 | 5.12E-05 | 3.12E-05 | 7.82E-04 | 5.02E-05 | 3.06E-05 |
| 7 | 9 | 5.30E-04 | 4.75E-05 | 3.09E-05 | 5.08E-04 | 4.67E-05 | 3.03E-05 |
| 8 | 9 | 3.65E-04 | 4.78E-05 | 3.11E-05 | 3.50E-04 | 4.70E-05 | 3.05E-05 |
| 5 | 10 | 1.37E-03 | 4.74E-05 | 2.48E-05 | 1.31E-03 | 4.65E-05 | 2.43E-05 |
| 6 | 10 | 8.16E-04 | 4.30E-05 | 2.44E-05 | 7.82E-04 | 4322E-05 | 2.39E-05 |
| 7 | 10 | 5.30E-04 | 3.92E-05 | 2.42E-05 | 5.08E-04 | 3.84E-05 | 2.37E-05 |
| 8 | 10 | 3.65E-04 | 3.64E-05 | 2.41E-05 | 3.50E-04 | 3.57E-05 | 2.36E-05 |
| 9 | 10 | 2.63E-04 | 3.66E-05 | 2.44E-05 | 2.53E-04 | 3.60E-05 | 2.39E-05 |
| 5 | 11 | 1.37E-03 | 4.03E-05 | 1.88E-05 | 1.31E-03 | 3.95E-05 | 1.84E-05 |
| 6 | 11 | 8.16E-04 | 3.66E-05 | 1.88E-05 | 7.82E-04 | 3.59E-05 | 1.84E-05 |
| 7 | 11 | 5.30E-04 | 3.32E-05 | 1.88E-05 | 5.08E-04 | 3.25E-05 | 1.84E-05 |
| 8 | 11 | 3.65E-04 | 3.03E-05 | 1.88E-05 | 3.50E-04 | 2.97E-05 | 1.84E-05 |
| 9 | 11 | 2.63E-04 | 2.83E-05 | 1.90E-05 | 2.53E-04 | 2.78E-05 | 1.86E-05 |
| 10 | 11 | 1.96E-04 | 2.85E-05 | 1.94E-05 | 1.88E-04 | 2.80E-05 | 1.90E-05 |
| 6 | 12 | 8.16E-04 | 3.13E-05 | 1.41E-05 | 7.82E-04 | 3.06E-05 | 1.38E-05 |
| 7 | 12 | 5.30E-04 | 2.84E-05 | 1.43E-05 | 5.08E-04 | 2.78E-05 | 1.40E-05 |
| 8 | 12 | 3.65E-04 | 2.58E-05 | 1.45E-05 | 3.50E-04 | 2.53E-05 | 1.43E-05 |
| 9 | 12 | 2.63E-04 | 2.37E-05 | 1.48E-05 | 2.53E-04 | 2.33E-05 | 1.45E-05 |
| 10 | 12 | 1.96E-04 | 2.23E-05 | 1.51E-05 | 1.88E-04 | 2.19E-05 | 1.48E-05 |
| 11 | 12 | 1.50E-04 | 2.26E-05 | 1.56E-05 | 1.44E-04 | 2.22E-05 | 1.53E-05 |

**Table 3.** Comparison between MC calculated $S$-values (Gy/Bq·s) of spherical concentric cells with $R_C = 5$ μm and $R_N = 4$ μm.

| | MC code | $S(N\leftarrow N)$ | *BrIccEmis*/MC code | $S(N\leftarrow Cy)$ | *BrIccEmis*/MC code | $S(N\leftarrow CS)$ | *BrIccEmis*/MC code |
|---|---|---|---|---|---|---|---|
| [67]Ga | *BrIccEmis*-condensed | 3.76E-03 | | 5.79E-04 | | 1.69E-04 | |
| | *BrIccEmis*-isolated | 3.71E-03 | | 5.78E-04 | | 1.69E-04 | |
| | MC4[a] | 3.62E-03 | +4% / +2% | 5.17E-04 | +12% / +12% | 1.41E-04 | +20% / +20% |
| | ETRACK[b] | 3.38E-03 | +11% / +10% | 5.81E-04 | 0% / -1% | 2.22E-04 | -24% / -24% |
| [99m]Tc | *BrIccEmis*-condensed | 1.54E-03 | | 9.08E-05 | | 5.43E-05 | |
| | *BrIccEmis*-isolated | 1.52E-03 | | 9.15E-05 | | 5.47E-05 | |
| | MC4[a] | 1.59E-03 | -3% / -4% | 8.46E-05 | +7% / +8% | 4.80E-05 | +13% / +14% |
| | GEANT4[c] | 1.57E-03 | -2% / -3% | 8.29E-05 | +10% / +10% | 4.99E-05 | +9% / +10% |
| | ETRACK[b] | 1.53E-03 | +1% / -1% | 7.41E-05 | +23% / +23% | 3.90E-05 | +39% / +40% |
| [111]In | *BrIccEmis*-condensed | 2.77E-03 | | 3.82E-04 | | 2.68E-04 | |
| | *BrIccEmis*-isolated | 2.70E-03 | | 3.83E-04 | | 2.69E-04 | |
| | MC4[a] | 2.85E-03 | -3% / -5% | 3.01E-04 | +27% / +27% | 2.64E-04 | +2% / +2% |
| | GEANT4[c] | 2.78E-03 | 0% / -3% | 3.60E-04 | +6% / +6% | 2.43E-04 | +10% / +10% |
| | ETRACK[b] | 2.72E-03 | +2% / -1% | 2.95E-04 | +29% / +30% | 1.88E-04 | +43% / +43% |
| [123]I | *BrIccEmis*-condensed | 2.86E-03 | | 3.06E-04 | | 1.94E-04 | |
| | *BrIccEmis*-isolated | 2.78E-03 | | 3.03E-04 | | 1.92E-04 | |
| | MC4[a] | 3.05E-03 | -6% / -9% | 3.35E-04 | -9% / -10% | 2.14E-04 | -9% / -10% |
| | ETRACK[b] | 2.93E-03 | -2% / -2% | 2.66E-04 | +15% / +15% | 1.42E-04 | +37% / +35% |
| [125]I | *BrIccEmis*-condensed | 6.51E-03 | | 6.44E-04 | | 3.72E-04 | |
| | *BrIccEmis*-isolated | 6.40E-03 | | 6.40E-04 | | 3.70E-04 | |
| | MC4[a] | 6.87E-03 | -5% / -7% | 6.94E-04 | -7% / -8% | 4.18E-04 | -11% / -11% |
| | ETRACK[b] | 6.78E-03 | -4% / -6% | 6.54E-04 | -2% / -2% | 3.61E-04 | +3% / +2% |
| | GEANT4[c] | 6.25E-03 | +4% / +2% | 6.37E-04 | +1% / 0% | 3.16E-04 | +18% / +17% |
| [201]Tl | *BrIccEmis*-condensed | 7.74E-03 | | 1.25E-03 | | 6.07E-04 | |
| | *BrIccEmis*-isolated | 7.14E-03 | | 1.24E-03 | | 6.05E-04 | |
| | MC4[a] | 8.26E-03 | -6% / -14% | 1.20E-03 | +4% / +4% | 5.51E-04 | +10% / +10% |
| | ETRACK[b] | 7.83E-03 | -1% / -9% | 1.27E-03 | -1% / -2% | 6.48E-04 | -6% / -7% |

[a]Data from (Bousis *et al.*, 2010)
[b]Data from (Ftániková and Böhm, 2000)
[c]Data from (Šefl *et al.*, 2015)

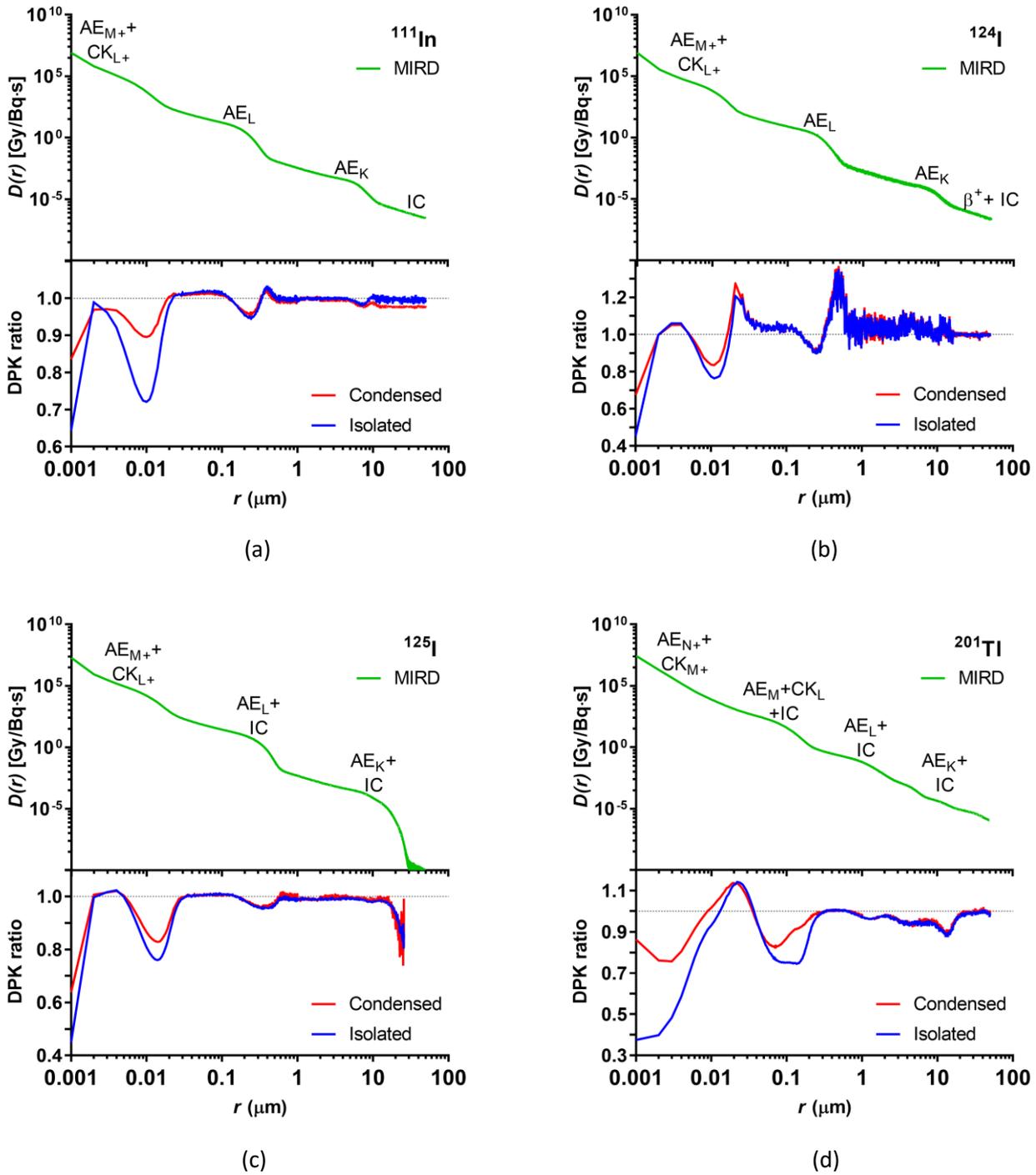

**Figure 1.** DPKs (Gy/Bq·s) for the AE+CK+IC spectra from MIRD-RADTABS (Eckerman and Endo, 2008a) of (a) $^{111}$In, (b) $^{124}$I, (c) $^{125}$I, and (d) $^{201}$Tl plotted on the top panel as a function of distance from a point source. DPK ratios, which are defined as DPK$_{BrIccEmis}$/DPK$_{MIRD\text{-}RADTABS}$, comparing dose distribution from MIRD-RADTABS data with spectra generated from *BrIccEmis* using condensed-phase and isolated-atom approximations plotted on the bottom panel.

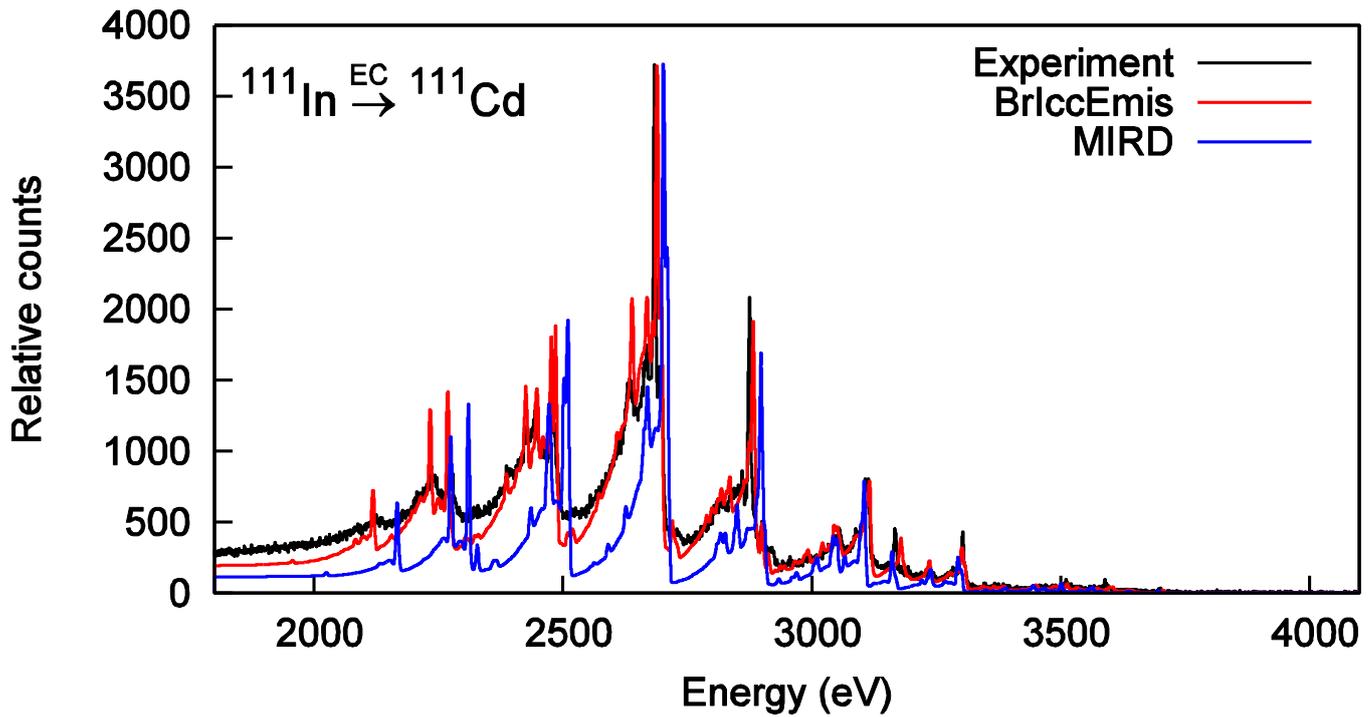

Figure 2. Comparison of Auger spectra from *BrIccEmis* and MIRD-RADTABS (Eckerman and Endo, 2008a) with experimental data (Yakushev *et al.*, 2005) within the energy range from 1800 to 3600 eV. There is a shift towards higher energies in the MIRD-RADTABS data. Both *BrIccEmis* and MIRD-RADTABS data were folded with the same approximate theoretical lineshape. Theoretical spectra were scaled to match the maximum intensity of the main L-Auger group at about 2700 eV.

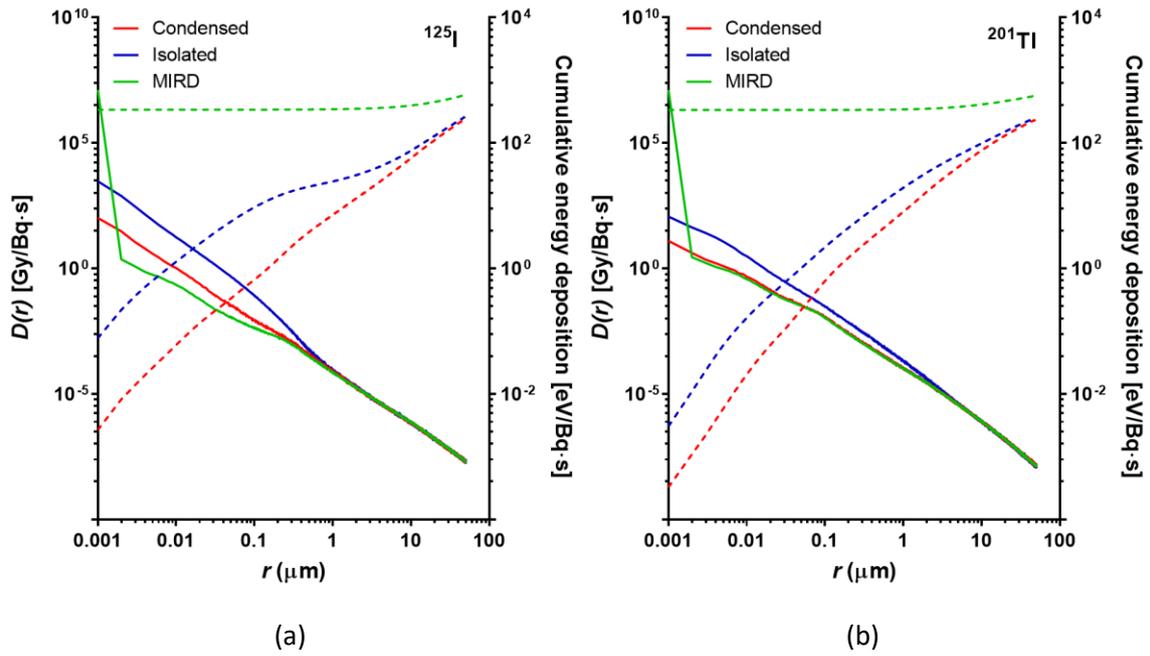

**Figure 3.** DPKs (Gy/Bq·s) for the X-ray spectra of (a) $^{125}$I and (b) $^{201}$Tl from *BrIccEmis* and MIRD-RADTABS (Eckerman and Endo, 2008a) plotted on the left y-axis as a function of distance from a point source (solid lines). Cumulated energy deposition (eV/Bq·s) as a function of distance from the point source is plotted on the right y-axis (dashed lines).

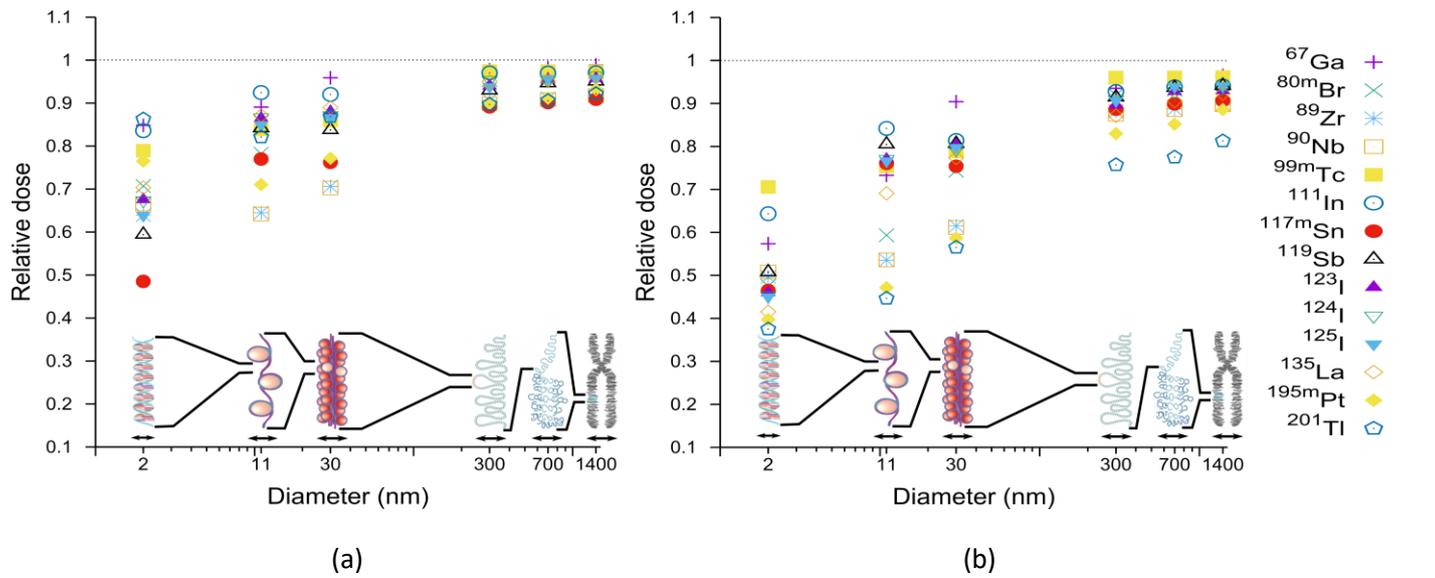

**Figure 4.** Relative dose, $S_{BrIccEmis}/S_{MIRD\text{-}RADTABS}$, comparing energy deposition based on MIRD-RADTABS data (Eckerman and Endo, 2008a) and *BrIccEmis* data calculated using condensed-phase (a) and isolated-atom (b) approximations. Relative dose is given as the ratio of energy deposited in spheres with diameters representing different DNA condensation states, i.e. DNA double helix (2 nm), DNA wrapped around histones (chromatin, 11 nm), chromatin fibre of packed nucleosomes (30 nm), chromosome section in extended form [300 nm], condensed section of chromosome (700 nm), and entire mitotic chromosome (1,400 nm).